\documentclass[letterpaper,prx,aps,twocolumn,showpacs,superscriptaddress,nofootinbib,longbibliography]{revtex4-1}
\usepackage{bm,color}
\usepackage[T1]{fontenc}
\usepackage[utf8]{inputenc}
\usepackage{latexsym}
\usepackage{graphicx} 
\usepackage{amsmath} 
\usepackage{amsfonts} 
\usepackage{braket}
\usepackage{subfigure}
\usepackage{lineno}

\def \be{{\bf e}}

\def \mHz{ {\mathrm{Hz} }}
\def \mD{ {\mathrm{D} }}
\def \mosc{ {\mathrm{osc} }}
\def \msite{ {\mathrm{site} }}
\def \mum{\mu \mathrm{m}}

\def \DD{{\Delta}}

	\newcommand{\bbm}{\begin{pmatrix}}
	\newcommand{\ebm}{\end{pmatrix}}

	\usepackage[normalem]{ulem} 
	\definecolor{mgreen}{RGB}{1,123,0}

\def\be{\begin{equation}}
\def\ee{\end{equation}}

\begin{document}
\title{Formation of spontaneous density-wave patterns \\in DC driven lattices}
\author{H. P. Zahn}
\affiliation{Institut f{\"{u}}r Laserphysik, Universit{\"{a}}t Hamburg, 22761 Hamburg, Germany}
\author{V. P. Singh}
\affiliation{Zentrum f{\"{u}}r Optische Quantentechnologien, Universit{\"{a}}t Hamburg, 22761 Hamburg, Germany}
\affiliation{Institut f{\"{u}}r Theoretische Physik, Leibniz Universit{\"{a}}t Hannover, Germany}
\author{M. N. Kosch}
\affiliation{Institut f{\"{u}}r Laserphysik, Universit{\"{a}}t Hamburg, 22761 Hamburg, Germany}
\author{L. Asteria}
\affiliation{Institut f{\"{u}}r Laserphysik, Universit{\"{a}}t Hamburg, 22761 Hamburg, Germany}
\author{L.~Freystatzky}
\affiliation{Zentrum f{\"{u}}r Optische Quantentechnologien, Universit{\"{a}}t Hamburg, 22761 Hamburg, Germany}
\affiliation{The Hamburg Centre for Ultrafast Imaging, 22761 Hamburg, Germany}
\author{K. Sengstock}
\affiliation{Institut f{\"{u}}r Laserphysik, Universit{\"{a}}t Hamburg, 22761 Hamburg, Germany}
\affiliation{Zentrum f{\"{u}}r Optische Quantentechnologien, Universit{\"{a}}t Hamburg, 22761 Hamburg, Germany}
\affiliation{The Hamburg Centre for Ultrafast Imaging, 22761 Hamburg, Germany}
\author{L. Mathey}
\affiliation{Institut f{\"{u}}r Laserphysik, Universit{\"{a}}t Hamburg, 22761 Hamburg, Germany}
\affiliation{Zentrum f{\"{u}}r Optische Quantentechnologien, Universit{\"{a}}t Hamburg, 22761 Hamburg, Germany}
\affiliation{The Hamburg Centre for Ultrafast Imaging, 22761 Hamburg, Germany}
\author{C. Weitenberg}
\email{christof.weitenberg@physnet.uni-hamburg.de}
\affiliation{Institut f{\"{u}}r Laserphysik, Universit{\"{a}}t Hamburg, 22761 Hamburg, Germany}
\affiliation{The Hamburg Centre for Ultrafast Imaging, 22761 Hamburg, Germany}


\begin{abstract}
    Driving a many-body system out of equilibrium induces phenomena such as the emergence and decay of transient states, which can manifest itself as pattern and domain formation. The understanding of these phenomena expands the scope of established thermodynamics into the out-of-equilibrium domain. Here, we experimentally and theoretically study the out-of-equilibrium dynamics of a bosonic lattice model subjected to a strong DC field, realized as ultracold atoms in a strongly tilted triangular optical lattice. We observe the emergence of pronounced density wave patterns -- which spontaneously break the underlying lattice symmetry -- using a novel single-shot imaging technique with two-dimensional single-site resolution in three-dimensional systems, which also resolves the domain structure. Our study suggests that the short-time dynamics arises from resonant pair tunneling processes within an effective description of the tilted Hubbard model. More broadly, we establish the far out-of-equilibrium regime of lattice models subjected to a strong DC field, as an exemplary and paradigmatic scenario for transient pattern formation.
\end{abstract}

\maketitle

	\begin{figure}[t]
		\includegraphics[width=0.95\linewidth]{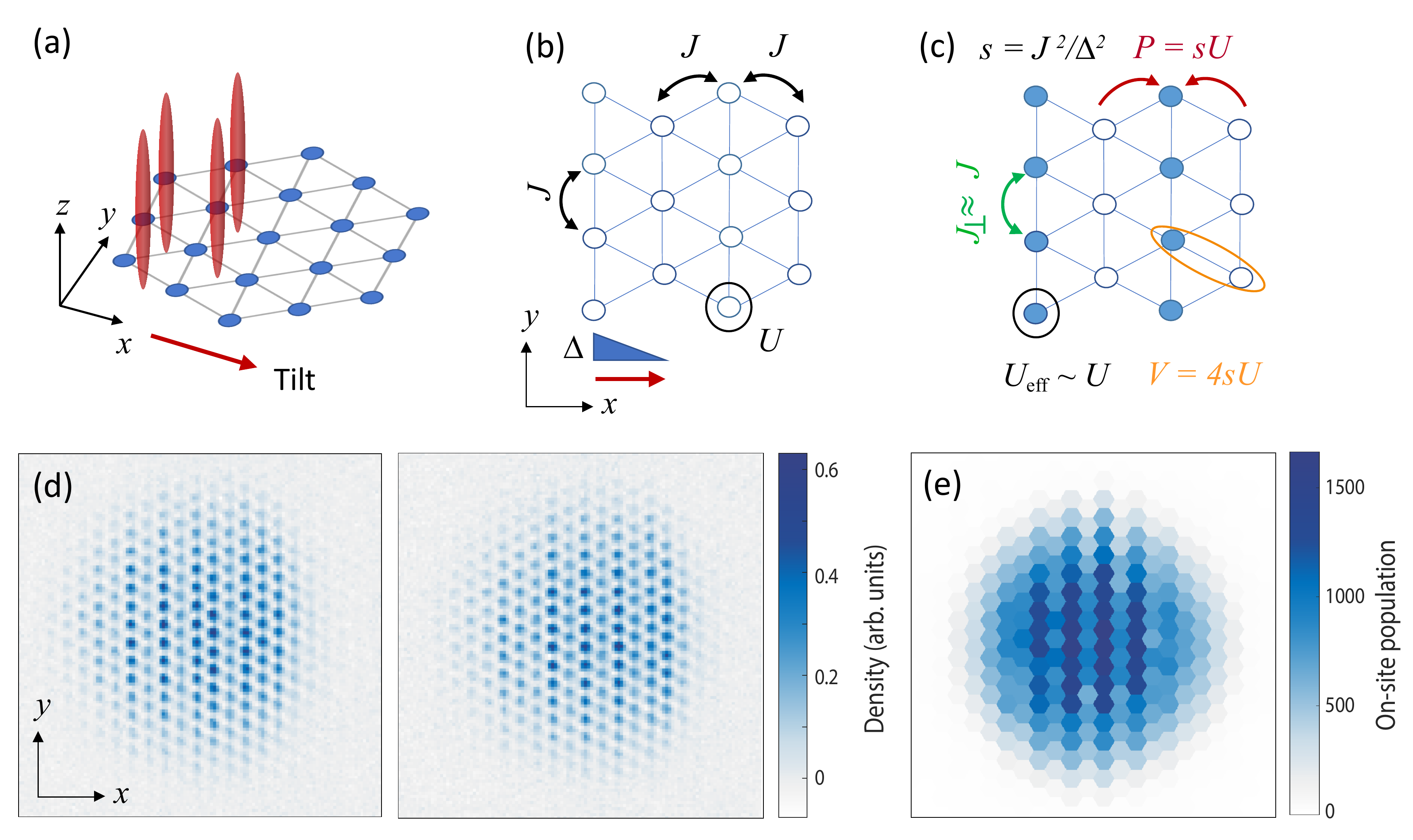}
		\caption{Bose-Einstein condensates in strongly-tilted optical lattices. (a), The system consists of a triangular lattice of tubes with a tilt applied perpendicularly to a lattice vector. (b), The system can be described by a Hubbard model with tunnel coupling $J$, on-site interaction $U$ and energy offset $\Delta$ along the tilt direction. (c), In the large tilt limit the system can be described by an effective Hamiltonian with no tilt and no single-particle tunneling in the tilt direction, but pair tunneling $P$ and nearest neighbour interaction $V$ with a strength set by $s U$ with $s=(J/\Delta)^2$. Tunneling perpendicular to the tilt $J_\perp$ and the effective on-site interaction $U_{\rm eff}$ remain approximately unchanged. (d), Experimental images of the spatial density show the spontaneous formation of charge density waves with wave-vector parallel to the tilt direction. The two example images are for identical parameters of a hold time of $60\,$ms and tilt with energy offset $\Delta=h \times 1.4\,$kHz, but one shows a domain wall in the center of the cloud and the other does not, directly reflecting the spontaneous nature of the pattern. The tunnel coupling is $J = h \times 13\,$Hz throughout the manuscript. (e), c-field simulation for the same parameters as in (d). The simulation does not model the density modulation within the lattice sites and the density distribution is therefore shown as tiled hexagons.}\label{fig:1_System}
	\end{figure} 

	\begin{figure}[t]
		\includegraphics[width=0.75\linewidth]{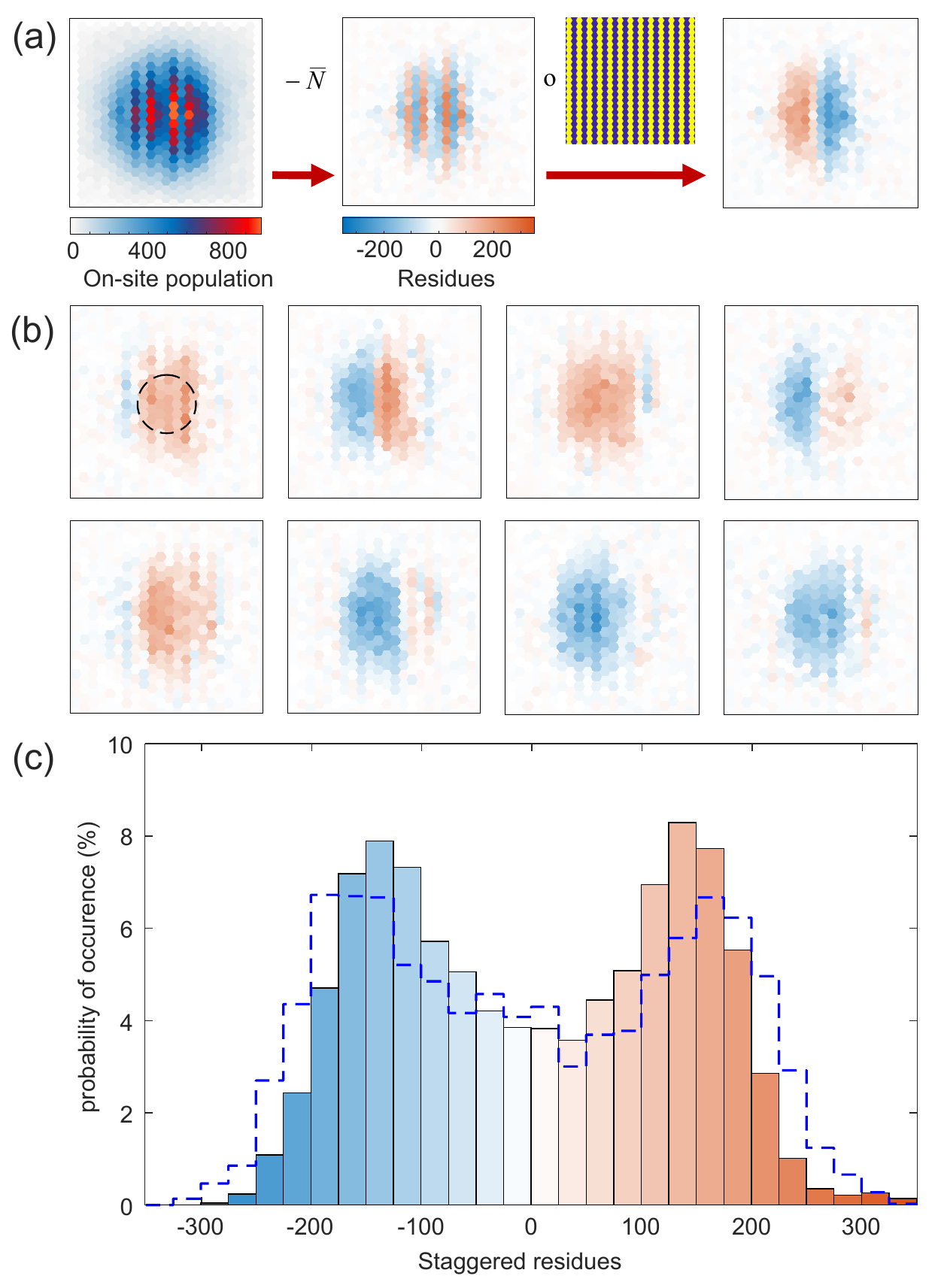}
		\caption{Spontaneous symmetry breaking. (a), The local density-wave order is quantified by calculating the residues after subtracting the mean density of about 130 identically prepared clouds. Multiplying by a staggering of fixed phase yields the staggered residues, which act as domain identifier.  (b), Staggered residues for typical individual images with $\Delta=h \times 1.4\,$kHz energy offset and $60\,$ms hold time illustrating the spontaneous formation of domains in the density-wave order with typically one or two large domains. (c), Histogram of the staggered residues for all lattice sites within a region of interest (dashed circle of radius 3 sites in (b)). The dashed line represents the histogram for a numerical simulation with $k_{\rm B} T/J = 250$ and a peak density of 900 atoms per tube. 
        }
		\label{fig:2_Domains}
	\end{figure}

\section{Introduction}

The appearance of patterns, which spontaneously break the spatial symmetry of the underlying potential landscapes, belongs to the most fascinating aspects in physics. Famous examples range from sand-dune-ripples to soliton trains in shallow water, where macroscopic patterns appear in driven systems on a homogeneous background landscape with continuous symmetry~\cite{Cross1993}. Such phenomena can be probed in the quantum regime using ultracold atoms, where pattern formation arises e.g. due to periodic driving \cite{Zhang2020} or due to long-range interactions as supersolid stripe pattern \cite{Bottcher2021}. Of special interest are periodic potentials, in which density waves or spin waves can spontaneously break the underlying lattice symmetry in a discrete fashion as observed, e.g. for tilted Bose-Hubbard models \cite{Sachdev2002,Pielawa2011,Simon2011,Meinert2014} or for spin chains formed by Rydberg-atom arrays \cite{Bernien2017}.

Here we observe the formation of a pronounced density wave for a Bose-Einstein condensate in a two-dimensional triangular lattice of tubes after applying a strong tilt, i.e. a spatially periodic modulation of the lattice occupations. The observed density wave can be attributed to the resonant pair tunneling and nearest-neighbour interaction terms that emerge as dominant terms in the effective Hamiltonian of the DC driven system and it appears on a time scale of $\sim 30\,$ms. At later times, the interplay of the direct tunneling and the emergent terms leads to an intriguing open-system dynamics, where the density-wave period increases. While density waves can appear as ground states for extended Hubbard models \cite{Dutta2015}, the ground state for our parameters has no density wave and the observed pattern formation occurs far from equilibrium after the quench into the tilted system. For a conceptual understanding of the early-time dynamics, we derive the effective Hamiltonian for bosons on the strongly-tilted triangular lattice and use a c-field simulation to both confirm the experimental observations and the scalings of the effective Hamiltonian.

We employ a novel microscopy technique \cite{Asteria2021}(Appendix B) as an indispensable tool to directly image the density-wave order and its domains in the 2D lattice of tubes with single-site resolution. This microscopy directly reveals the spontaneous symmetry breaking in the pattern formation and domains thereof. Our striking observation of spontaneous pattern formation is very distinct from recent experiments in tilted optical lattices in the regime of unit occupation \cite{Guardado-Sanchez2020,Scherg2021}, where a suppressed decay of initially prepared density waves due to Stark many-body localization or kinetic constraints, but no pattern formation was observed. In contrast, occupation of period-doubled states was previously inferred from momentum-space images in the regime of large bosonic filling factors in an array of tubes, similar to our system that we resolve in real-space with the novel microscopy technique. However, this pattern formation was triggered by the different mechanism of dynamical instability in fully coherent systems, which occurs both in moving lattices \cite{Fallani2004, Cristiani2004} and in periodically driven lattices \cite{Gemelke2005}.

\section{Experimental Setup}

Our experiments start with a Bose-Einstein condensate (BEC) of $\sim 5 \times 10^4$ $^{87}$Rb atoms in a triangular optical lattice with tunnel coupling $J=h \times 13\,$Hz and transverse confinement $\omega_z=2\pi \times 30\,$Hz, for which we estimate an equivalent Hubbard interaction term $U\sim h \times 2.3\,$Hz (Fig.~\ref{fig:1_System}a). The lattice laser wavelength is $\lambda=1064\,$nm yielding a lattice constant $a=2\lambda/3$ and a distance between lattice sites in the tilt direction of $\tilde{a}=a \sqrt{3}/2$. We apply a tilt by shifting the magnetic trap, let the system evolve and image the real space distribution with the quantum gas magnifier \cite{Asteria2021}, i.e. by magnifying the distribution using matter wave optics prior to optical imaging. 

We identify the relevant regime of tilts, where the atoms are Stark localized and form a metastable system on the slope of the potential \cite{Burger2001, Oppong2020}, while still showing sufficiently strong pair tunneling driving the density-wave formation. We work at a tilt of $F= h \times 2.3\,$kHz/µm, which leads to an energy offset between neighboring lattice sites of $\Delta=F \tilde{a}= h \times 1.4\,$kHz and to a large suppression of tunneling $J/\Delta=0.009$. The resulting density distributions for various hold times are presented in Fig.~\ref{fig:1_System}d and show the pronounced density waves with random phase constituting spontaneous symmetry breaking as we discuss below. We model the system with extensive c-field simulations (Appendix A) and find qualitative agreement when comparing the site occupations (Fig.~\ref{fig:1_System}e) as also discussed below. 

\section{Theoretical Modelling}

To describe the system, we introduce an effective Hamiltonian (Appendix A): the energy offset $\Delta$ introduces a relative phase evolution, which can be treated in a Floquet picture yielding an effective Hamiltonian $\hat{H}_{\rm eff}$ in a high-frequency approximation. $\hat{H}_{\rm eff}$ contains new processes with a scale $s=(J/\Delta)^2$: a center-of-mass-conserving pair tunneling, where one atom goes up the tilt and the other goes down, with strength $P=s U$, and a next neighbor interaction with strength $V=4 s U$ (Fig.~\ref{fig:1_System}b,c). These processes are dominant, because the direct tunneling $J$ along the tilt is suppressed and vanishes in $\hat{H}_{\rm eff}$. The choice of the triangular lattice enhances the pair tunneling dynamics, because each lattice site has two neighbours on top and below, which gives rise to four possible pair tunneling terms. This Hamiltonian features a density-wave as ground state in a regime of larger $P$ and $V$ (see Monte Carlo studies in the Supplementary Information \cite{SupMat}), whereas here, interestingly, the density wave appears as a transient non-equilibrium phenomenon after the quench and for this dynamics, the pair tunneling $P$ sets the relevant scale. This observation also suggests an interpretation as a prethermalized state \cite{Ueda2020} of the time evolution triggered by the quench, in which the system does not directly relax to a thermal state of either the underlying or the effective Hamiltonian, but rather relaxes in a two-step process via a density-ordered state.

	\begin{figure}[t!]
		\includegraphics[width=0.8\linewidth]{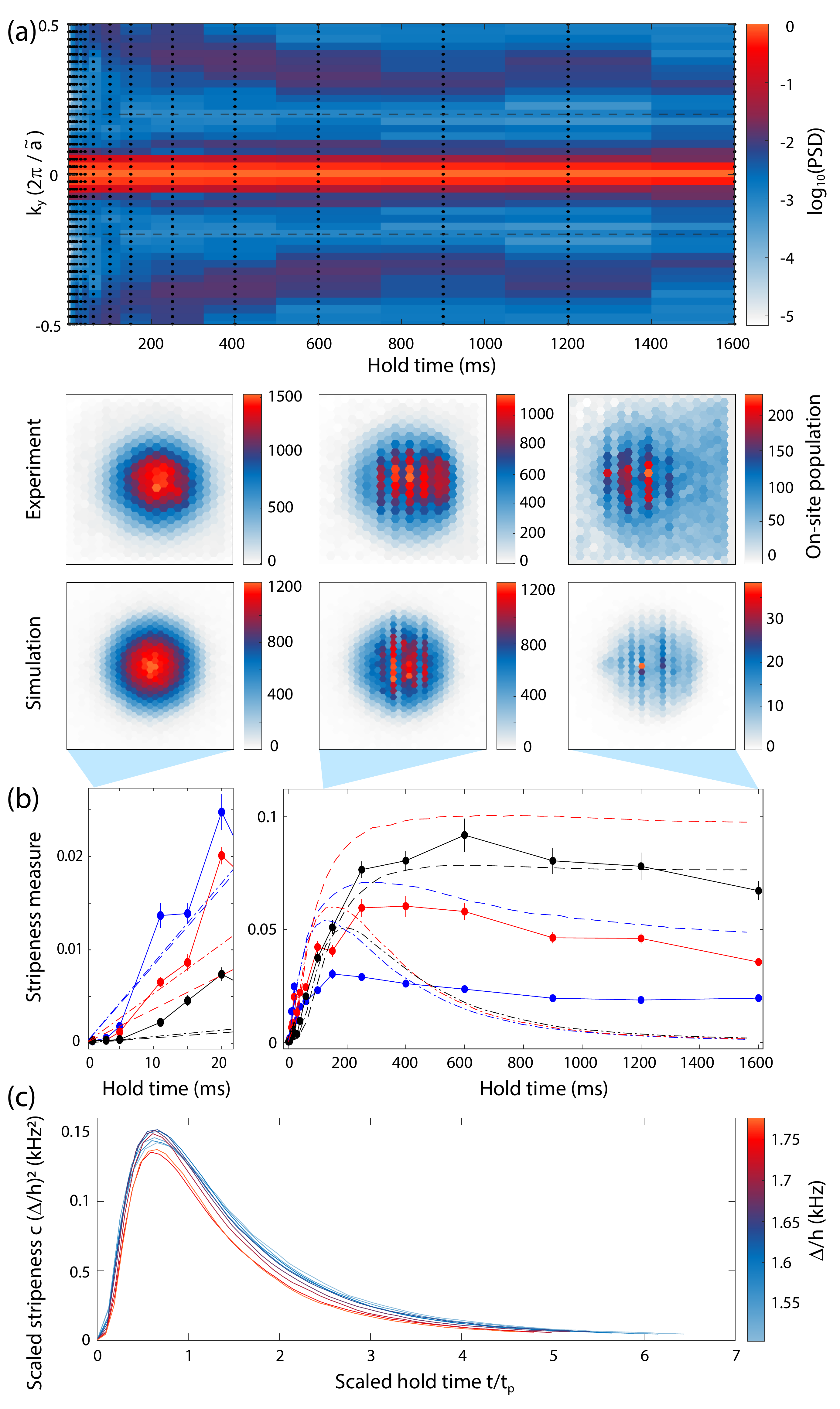}
		\caption{Formation dynamics and decay of the density wave. (a), Normalized power spectral density of the column occupations for a tilt with energy offset $\Delta = h \times 1.4$\,kHz initially develops a peak at $k\tilde{a}/(2\pi) = \pm1/2$ (period 2), which splits and moves to $k\tilde{a}/(2\pi) \sim \pm1/3$ (period 3) at longer time. The dots show the exact parameters for which data was taken. The dashed lines mark $k\tilde{a}/(2\pi) = \pm1/5$ which is used in the computation of the stripeness measure (see main text). (b), Dynamics of the stripeness measure $c$ for different evolution times after the quench into the tilted system showing the fast build-up and slow decay of the patterns. The color encodes the tilt strength of $\Delta =h \times 1.1$\,kHz (blue), $\Delta = h \times 1.4$\,kHz (red), $\Delta = h \times 1.7$\,kHz (black). The experimental data (circles) is compared to a numerical c-field simulation including the atom number loss (dashed lines), and without loss (dash-dotted lines). The numerical simulation without atom loss shows a faster decay. The simulation uses a temperature of $k_{\rm B} T/J =100$. The insets show example single-shot density distributions from the experiment (top row, hold time = 1\,ms, 60\,ms, 1600\,ms) and the simulation with atom loss (bottom row, hold time = 0\,ms, 64\,ms, 1560\,ms). All error bars correspond to the $68\%$ confidence interval. (c), The density-wave order of the simulated data without atom loss follows the scaling of the effective Hamiltonian and collapses on a curve when the time axis is rescaled by the pair tunneling time $t_p$ and the density-wave order is scaled by the energy offset squared $\Delta^2$.} \label{fig:3_Formation}
	\end{figure} 

\section{Spontaneous Symmetry Breaking}

The density-wave has twice the period of the underlying lattice and can therefore spontaneously choose between two positions, giving rise to a discrete symmetry breaking. We quantify this by a density-wave order of positive or negative sign, depending on whether the density wave is in or out of phase with a globally fixed reference pattern. This is realized by multiplying the residues with this reference pattern of alternating positive and negative sign (Fig.~\ref{fig:2_Domains}a). We find that this density-wave order forms large domains of a random sign of the order (Fig.~\ref{fig:2_Domains}b), demonstrating the spontaneous nature of the symmetry breaking. The double-peak structure of a histogram of the local density-wave order values reflects the spontaneous symmetry breaking as well (Fig.~\ref{fig:2_Domains}c). This is further supported by the persistence of the double-peak structure after post selection on the position of the lattice relative to the initial cloud center \cite{SupMat}, which is not triggering the symmetry breaking. For longer hold times we observe a pinning of the density wave, which we discuss in the Supplementary Information \cite{SupMat}. 

\section{Emergence and Decay of Density-Wave Order}

Moving further, we study the intriguing dynamics of the density-wave order, which is characterized by a spontaneous emergence and a very long-lived decay. First, we quantify the period of the density wave pattern via the power spectral density (PSD) $|n_k|^2 = |(\mathcal{F}n_x)(k)|^2 / N_{\rm tot}^2$ of the column populations $n_x$ normalized by the total atom number $N_{\rm tot}$, which captures the spectrum of the density-density correlations (Fig.~\ref{fig:3_Formation}a): we find an initial peak at $k\tilde{a}/(2\pi) = \pm1/2$ (period 2), which splits and moves to $k\tilde{a}/(2\pi) \sim \pm1/3$ (period 3) at longer times when the single-particle tunneling starts a transport down the slope. The appearance of the longer period during the intriguing dynamics at longer times remains to be explained.

We define as our stripeness measure the relative integrated PSD in this peak 
$ c = \int_{1/5}^{1/2} |n_k|^2 {\rm d}(k\tilde{a}/2\pi) / \int_{0}^{1/2} |n_k|^2 {\rm d}(k\tilde{a}/2\pi)$. This signal appears within $\sim 30\,$ms, reaches a maximum at around $\sim 200\,$ms, and then shows almost no sign of decay for over 1.6\,s (Fig.~\ref{fig:3_Formation}b). The long lifetime is captured well by a c-field simulation, when the atom number loss of the experiment is added to the simulation (Appendix A). This behavior points to the interaction-driven nature of the density-wave dynamics, which is frozen when the density drops \cite{SupMat}. 

For the understanding of the dynamics, it is instructive to refer to the effective Hamiltonian introduced above. It defines a time scale (Appendix A) $t_{\rm p}=h/(16 N P)\approx 210$\,ms for the density-wave formation dynamics, where we used $\Delta = h \times 1.4$\,kHz and where $N\approx 1500$ is the number of atoms in the central tube for the data in Fig.~\ref{fig:3_Formation}. The scaling $t_{\rm p} \propto \Delta^2$ fits well with the density-wave dynamics in the c-field simulation without atom number loss (Fig.~\ref{fig:3_Formation}c), which therefore confirms this Floquet picture. The dependence of $t_{\rm p}$ on $N$ explains why the dynamics is frozen with atom loss leading to long-lived density-wave patterns and why the simulation needs to include the atom number loss to reproduce the slow decay. We also study the influence of coherence between the lattice sites by realizing different temperatures and find that initial coherence is required for the formation of the density wave \cite{SupMat}.

	\begin{figure}[t]
		\includegraphics[width=0.8\linewidth]{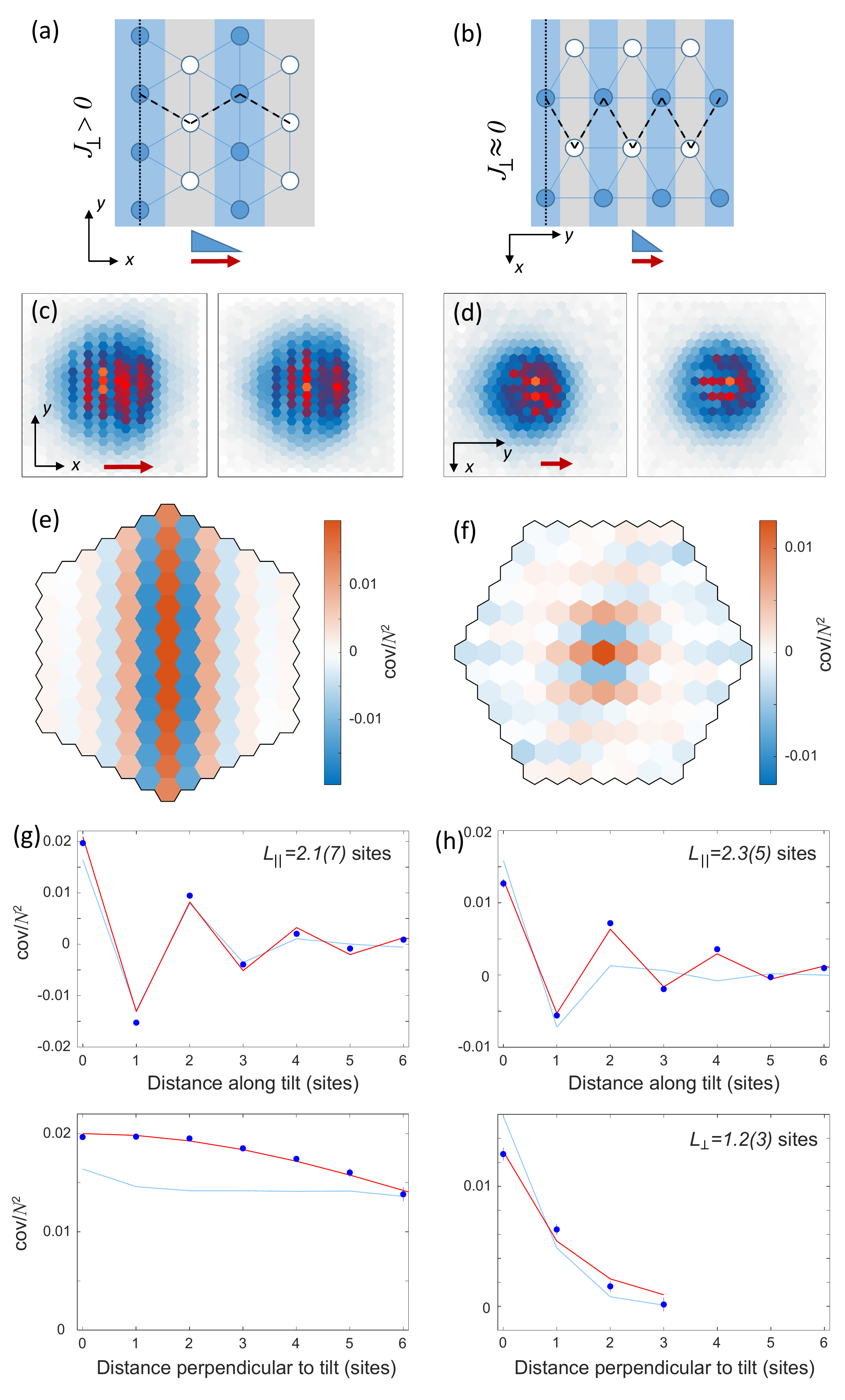}
		\caption{Dependence on the tilt direction. (a), Sketch of the triangular lattice geometry and the columns (grey and blue bars) perpendicular to the tilt for the two situations with different values of transverse tunneling $J_\perp$. (b), Sketch of the situation with the tilt in the perpendicular direction. Here tunneling within the columns only appears as a higher order process. For an easier comparison, we have rotated the system, such that the tilt always points to the right. (c), (d) Typical experimental images for the two tilt directions with energy offset $\Delta = h \times 1.4$\,kHz, a hold time of 60\,ms and a maximum atom number per tube of 900 and 850, respectively. The orientation of the images and the tilt direction (red arrow) are as in (a) and (b), respectively. (e), (f), The density correlation functions (Appendix C) for the two cases. (g), (h), Cuts of the density correlations functions show the density-wave order measured along a zig-zag path along the tilt direction (dashed line in (a) and (a)) and a monotonic decay within the columns (dotted line in (a) and (b)). The red lines show heuristic fits (Appendix C) to the data, the light blue lines result from the same analysis on pictures from numerical simulations with $k_{\rm B} T/J =50$. All error bars correspond to the $68\%$ confidence interval.} \label{fig:4_TiltDirection}
	\end{figure} 

\section{Effect of Transverse Tunneling}

In another set of experiments we study the effect of transverse tunneling, i.e. tunneling perpendicular to the tilt. We compare the situation of two different tilt directions, which have equal tight-binding models except for a very different strength of the transverse tunneling within the triangular lattice (Fig.~\ref{fig:4_TiltDirection}a,b). The resulting images reveal that the density-wave pattern becomes more irregular without the transverse tunneling (Fig.~\ref{fig:4_TiltDirection}c,d), which is directly reflected in the density correlations (Fig.~\ref{fig:4_TiltDirection}e,f). In both cases, we find staggered density correlations of the density-wave order along the tilt and monotonically decaying correlations perpendicular to the tilt, but with different ranges (Fig.~\ref{fig:4_TiltDirection}g,h). We heuristically fit the decay and state the decay lengths $L_{\perp}$ and $L_{\parallel}$ in units of the number of lattice sites (Appendix C). Interestingly, the range of the staggered density-wave order along the tilt has the same decay length $L_{\parallel}$ of 2.1(7) and 2.3(5) sites for the strong and weak transverse tunneling, respectively. The correlation length perpendicular to the tilt $L_{\perp}$, however, is drastically reduced from long-range order to short range order with decay length $L_{\perp}$ of 1.2(3) sites when changing the tilt direction. 
The analysis shows that strong transverse tunneling aligns the phase of the symmetry breaking density-waves across the system while the range of the staggered order along the tilt is not affected.

\section{Conclusion}

In conclusion, we have observed the appearance of long-lived density-wave patterns for a BEC in a strongly-tilted two-dimensional optical lattice and confirmed their spontaneous breaking of the discrete lattice symmetry. The dynamics is supported by a Floquet picture, which introduces correlated pair tunneling processes and sets the density-dependent time scale. The findings continuously connect to the charge density wave observed in a tilted bosonic Mott insulator in the unit-filling regime \cite{Sachdev2002,Pielawa2011,Simon2011,Meinert2014}, with its mapping to quantum spin models. By going away from solvable models towards full-fledged high-dimensional many-body systems, we identify an intriguing regime of many-body dynamics.

With these experiments we establish the quantum gas magnifier for imaging spontaneously formed pattern on the scale of single lattice sites in 3D systems. The technique can in principle also image the coherence properties with high resolution \cite{Asteria2021} and in the future, it could be used to observe other exotic states such as the twisted superfluid \cite{Luhmann2016} or to resolve domains of spontaneous symmetry breaking of phase patterns in driven lattices \cite{Struck2013,Parker2013} and in higher bands \cite{Kock2015}. With the observation of spatial symmetry breaking in a non-equilibrium situation, we set the stage for observing density wave ground states by engineering stronger non-standard Hubbard terms \cite{Baier2016, Meinert2016}.

The description of the observed dynamics by the terms of a Floquet picture points a way to use DC driven systems \cite{Dimitrova2020,Bohrdt2021,Zisling2021} as opposed to AC driven systems for Floquet engineering of relevant extended Hubbard models \cite{Weitenberg2021}. While for AC driven Floquet systems many-body localization has been proposed for stabilization against heating, DC driven systems require Stark localization (and potentially many-body Stark localization \cite{Schulz2019}) for stabilization and our experiments show that it is promising to identify suitable regimes. Of particular interest would be extended Hubbard models in the strongly-correlated regime featuring exotic states such as topological Mott insulators \cite{Rachel2018} or other density-assisted tunneling terms for engineering artificial magnetic fields \cite{Jamotte2021}. Furthermore, the DC drive could be combined with an AC drive leading to rich phenomena such as topological bandgap solitons \cite{Mukherjee2020} or Stark time crystals \cite{Kshetrimayum2020}.
 
\section*{Acknowledgments} 
The work is funded by the Cluster of Excellence 'CUI: Advanced Imaging of Matter' of the Deutsche Forschungsgemeinschaft (DFG) - EXC 2056 - project ID 390715994, by the DFG Collaborative Research Center SFB 925, project ID 170620586, and by the DFG Research Unit FOR 2414, project ID 277974659. C.W. acknowledges funding by the European Research Council (ERC) under the European Union's Horizon 2020 research and innovation programme under grant agreement No. 802701. V.P.S. acknowledges funding by the Cluster of Excellence 'QuantumFrontiers' - EXC 2123 - project ID 390837967.
 

\section*{Appendix A: Numerical and analytic methods} 
\subsection*{c-field simulation} 
We simulate the dynamics of tube condensates arranged in a 2D triangular geometry using the classical-field method of ref.~\cite{Singh2016}. The system is described by the Bose-Hubbard model
\begin{equation}\label{eq:BH}
\hat{H} =  -\sum_{\braket{i j} } J_{ij}  (\hat{\psi}_i^\dagger \hat{\psi}_j  +  \hat{\psi}_i \hat{\psi}_j^\dagger   ) + \frac{\tilde{U}}{2} \sum_i \hat{n}_i^2 + \sum_i V_i \hat{n}_i,
\end{equation}
where $\hat{\psi}_i$ $(\hat{\psi}_i^\dagger)$ is the bosonic annihilation (creation) operator and $\hat{n}_i =  \hat{\psi}_i^\dagger \hat{\psi}_i$ is the occupation operator at site $i$.  $\braket{ij}$ denotes nearest-neighbor bonds and $J_{ij} $ are the corresponding tunneling energies. $\tilde{U}$ is the on-site interaction energy and  $V_i$ is the harmonic trap potential. 
The lattice site locations $\mathbf{r}_\mathbf{i} = (x_\mathbf{i}, y_\mathbf{i}, z)$ represent a 3D system, which is continuous in the $z$ direction. We choose the primitive vectors $\mathbf{a}_1 = a \mathbf{e}_y$ and $\mathbf{a}_2 = \sqrt{3}/2 \mathbf{e}_x + 1/2 \mathbf{e}_y$
resulting in $\mathbf{r}_\mathbf{i} = i_1 \mathbf{a}_1 + i_2 \mathbf{a}_2 + z \mathbf{e}_z$, with $a = 709\,$nm being the lattice constant.
We have the same tunneling energy $J = h \times 13 \, \mHz$ for all 6 neighbors. 
The 3D repulsive interaction is determined by $g_{3\mD} =  h^2 a_s/(\pi m)$, where $a_s$ is the $s$-wave scattering length and $m$ is the atomic mass. For the tubes, this results in an effective 1D interaction $g_{1\mD} = g_{3\mD}/(2 \pi a_\mosc^2)$, with $a_\mosc$ being the oscillator length determined by the lattice depth and the recoil energy. 
Using the experimental parameters we obtain $g_{1\mD}= 7.06 J \mum$.
For numerical simulations we discretize the $z$ direction with a discretization length of $\ell_z =0.4\, \mum$. 
This introduces an additional effective coupling in $z$ direction  $J_z = h^2/(8 \pi^2 m \ell_z^2) = 27.9 J$ and the on-site interaction is rescaled to $U_\msite = g_{1\mD}/\ell_z = 17.6 J$ (ref.~\cite{Mora2003, Struck2013}). We use the same trap frequencies $\omega_{x, y} = 2 \pi \times 135\, \mHz $  and $\omega_z  = 2 \pi \times 30\, \mHz$ as in the experiments. The atom number of the central tube is chosen according to the experimental values. The temperature is more difficult to determine in the experiment and the temperature in the numerics is chosen to match the experimental results as stated in the respective captions.
 
We simulate this system using a numerical lattice of size $N_x \times N_y \times N_z$, where $N_x$ and $N_y$ are chosen between $50-100$ and $N_z$ is fixed at $N_z=81$. In our c-field representation we replace the operators $\hat{\psi}$ in Eq.~(\ref{eq:BH}) and in the equations of motion by complex numbers $\psi$. We sample the initial states in a grand-canonical ensemble of temperature $T$ and chemical potential $\mu$ via a classical Metropolis algorithm. We then propagate each state using the classical equations of motion. We calculate the observable such as the tube density $n(x, y, t)$ and average over the initial ensemble. For tilting the system, we adopt the experimental procedure and therefore abruptly displace the harmonic trap by a displacement $d$ either along the lattice direction or perpendicular to the lattice direction. To implement atom loss, we additionally include the term $-\gamma n_i$ in the equations of motion, where $\gamma$ is the loss rate. We use the experimentally determined $\gamma$ from the time evolution of density up to $100\, \mathrm{ms}$. We note that the measured $\gamma$ varies depending on the value of $d$. 

\subsection*{Derivation of the Hubbard interaction strength} 
We estimate an effective Hubbard interaction strength $U$ using the Ansatz for the wavefunction in a single lattice site $\varphi(x,y,z)=\varphi_{2D}(x,y)\varphi_z(z)$, where $\varphi_{2D}(x,y)$ is the wavefunction in the lattice plane and it is determined by the lattice depth, and $\varphi_z(z)$ is the wavefunction in $z$ direction which we assume to be a Thomas-Fermi profile. We determine $\varphi(x,y,z)$ for $N=1000$ atoms and extract the total interaction energy as $E_{\rm int}=N^2 g_{3\mD} \int|\varphi(x,y,z)|^4 {\rm d}x {\rm d}y {\rm d}z$. We obtain an effective Hubbard strength $U=\frac{2}{N^2}E_{\rm int}$ of $\sim h \times 2.3\,$Hz. This derivation neglects the change of the Thomas Fermi profile with the atom number per tube, which is valid for small relative atom number changes during the initial dynamics. This effective Hubbard $U$ is only used for the computation of $t_p$ via the effective Hamiltonian, but not for the description via the c-field simulation, where a different Hubbard $U_\msite$ appears for the description with discretized $z$ direction.

\subsection*{Effective Hamiltonian of the tilted lattice} 
We map the DC driven lattice Hamiltonian to a periodically driven one in the interaction picture and use Magnus expansion to second order to derive an effective Hamiltonian. The resulting Hamiltonian has the form (see full derivation in the Supplementary Information \cite{SupMat})
\begin{equation}\label{eq:Heff}
	\hat{H}_{\rm eff}=\hat{H}_{U,\rm eff}+\hat{H}_{\rm P}+\hat{H}_{J_\perp},
\end{equation}
$\hat{H}_{U,\rm eff}$ describes the effective interaction with nearest neighbor interaction of strength $V=4sU$ and renormalized on-site interaction of strength $U_{\rm eff}=U-8sU$, where we used the scaling factor $s=J^2/\DD^2$. $\hat{H}_{\rm P}$ denotes the pair tunneling with strength $P=sU$ and $\hat{H}_{J_\perp}$ is the term describing tunneling perpendicular to the tilt. These terms are
\begin{align}
	\hat{H}_{U,\rm eff} &=\frac{U_{\rm eff}}{2}\sum_j\hat{n}_j(\hat{n}_j-1) \, \nonumber \\
	& \quad +V\sum_{\langle jk\rangle_y}\hat{n}_j\hat{n}_k, \\
	\hat{H}_{\rm P} &=-P\sum_{\langle jkl\rangle_\times}\hat{b}_j^\dagger \hat{b}_j^\dagger \hat{b}_k \hat{b}_l+\hat{b}_l^\dagger \hat{b}_k^\dagger \hat{b}_j \hat{b}_j,\\
	\hat{H}_{J_\perp} &=-\sum_{\langle jk\rangle_x}(J\hat{b}_j^\dagger \hat{b}_k + sU\hat{b}_j^\dagger \hat{n}_j \hat{b}_k + sU\hat{b}_j^\dagger \hat{n}_k \hat{b}_k +h.c.) \nonumber\\
	&\quad +2sU\sum_{\langle jkl\rangle_\triangle}(\hat{b}_j^\dagger \hat{n}_l \hat{b}_k + \hat{b}_k^\dagger \hat{n}_l \hat{b}_j).
\end{align}
$\hat{b}_j$ is the bosonic annihilation operator on lattice site $j$ and $\hat{n}_j=\hat{b}_j^\dagger \hat{b}_j$. $ \langle jk \rangle_{x(y)}$ denotes nearest neighbours in the $x(y)$ direction, $\langle jkl\rangle_\times$ are all pairs of bonds with one bond up and one down the tilt, and $\langle jkl\rangle_\triangle$ denotes triangular plaquettes such that $j$ and $k$ are nearest neighbors in the $y$ direction.
In particular, this derivation establishes the energy scale $P=J^2 U /\Delta^2$ of the pair tunneling terms, which is relevant for the density-wave formation. 
The full derivation and the Hamiltonian for the other tilt direction are given in the Supplementary Information \cite{SupMat}.

\subsection*{Explanation of $t_{\rm p}$} 
From the pair tunneling we estimate a time scale of $t_{\rm p}=h/(16 N P)$ until the maximal amplitude of the density wave, where $P$ is the strength of the pair tunneling derived above. Compared to a standard two-level system with coupling $\Omega$ reaching maximal transfer after $h/(4\Omega)$ we added an additional factor of $4N$ here, where 4 is the number of possible pair tunneling processes in the tilted triangular lattice and $N$ accounts for the density dependence of the process in mean field taking into account the bosonic enhancement.

\section*{Appendix B: Experimental setup and methods}
\subsection*{Optical lattice setup} 
Our optical triangular lattice setup consists of three linearly polarized running waves \cite{Becker2010, Flaschner2016} of wave vector ${\bf k}_i$ with $\lvert {\bf k}_i \rvert=2\pi/\lambda$ intersecting under an angle of $120^\circ$. The resulting potential can be written as

\begin{equation}
	V_{\rm lat 2D}({\bf r}) = -2 V_{\rm lat} \sum_{i>j} \cos\left(\left({\bf k}_i - {\bf k}_j\right){\bf r} \right) 
\end{equation}

where the lattice depth $V_{\rm lat}$ is proportional to the intensity of the lattice beams. We neglected the phases of the beams with respect to each other because they only result in a global shift of the lattice.

\subsection*{Site-resolved imaging} 
We use the technique introduced in ref.\cite{Asteria2021} to magnify the density distributions prior to imaging. We freeze the density distribution in a deep lattice of $V_{\rm lat} \sim 6E_{\rm r}$ with $E_{\rm r} = h^2/(2m\lambda^2)$, $m = 87\,$u being the mass of ${}^{87}$Rb, resulting in a tunnel coupling of $J \sim h \times 10^{-3}\,$Hz. This allows to ramp to a magnetic confinement of $\omega_{\rm pulse} \sim 2\pi \times 300\,$Hz which in combination with a subsequent time-of-flight $t_{\rm tof} \sim 20\,$ms yields typical matter wave magnification of $\sim 40$. We extract the on-site populations by fitting a triangular lattice to the data and subsequently summing over the Wigner-Seitz cells around the individual sites. For the images with tilt, we shift back the magnetic trap center for the matter-wave magnification, in order to reduce anharmonic matter-wave aberrations.

\subsection*{Parameter calibration} 
We calibrate the lattice depth $V_{\rm lat}$ via Kapitza-Dirac scattering and determine the corresponding tunneling element $J$ via a band structure calculation. The atom loss rate is determined to be $\gamma = (2.3(2), 2.9(2), 3.6(2))\,{\rm s}^{-1}$ for energy offsets $\Delta =h \times (1.1, 1.4, 1.7)$\,kHz  from an exponential fit to the first $100\,$ms of the experiment. Note that the atom loss measures the number of atoms that get lost from the slope, thus it includes atoms that fall to the trap center, which is the reason for the tilt dependence. The experiments are performed with $^{87}$Rb atoms in the $F=2$, $m_F=2$ state in a Ioffe-Pritchard-type magnetic trap with a coil winding pattern in between a cloverleaf and a 4D configuration and with trapping frequencies $\omega_{x,y}= 2\pi \times 135\,$Hz in plane calibrated via an oscillation and $\omega_z=2\pi \times 30\,$Hz transversally, in the latter direction dominated by the optical lattice beams. The tilt is produced by shifting the position of the magnetic trap using additional magnetic field coils. We determine the tilt from the measured shift distance of typically $d=15\,$µm and cross check with the period of Bloch oscillations induced by the tilt. The variation of the tilt over 10 lattice sites resulting from the unchanged harmonic confinement is around $\pm 20\%$ and is also included in the c-field simulation by modeling the full trap.

\section*{Appendix C: Data evaluation}
\subsection*{Residue computation} 
To compute the residues $\delta N_{jk}^{(i)}$ at site $(j,k)$ we proceed for every single shot $i$ in the following way: We average all images corresponding to the same experimental parameters and read out the lattice site populations of this mean image with the same masks as the ones used for the shot we want to compute the residues of. The masks of different shots differ by the phase of the optical lattice (see discussion of the lattice-phase drifts in the Supplementary Material \cite{SupMat}). The resulting populations are used as reference and are subtracted from the on-site populations of the single shot.

\subsection*{Correlation computation} 
The first step in the evaluation is to compute the covariance ${\rm cov}_{jkj'k'} = 1/(N-1) \sum_{i=1}^N \delta N_{jk}^{(i)} \delta N_{j'k'}^{(i)}$ of the residues. This is done for every site within 3 sites of the center with respect to every other site satisfying the same condition. Subsequently the covariance is averaged for pairs of points having the same or exactly opposite distance vector and is eventually normalized to the mean maximal tube population for the corresponding parameters. 
The error of the covariance is computed as $1/\sqrt{N} \sqrt{\sum_{i=1}^N ({\rm cov}_{jkj'k'}^{(i)} - {\rm cov}_{jkj'k'})^2}$ where ${\rm cov}_{jkj'k'}^{(i)} = \delta N_{jk}^{(i)} \delta N_{j'k'}^{(i)}$. This error is then propagated when computing the mean. 

\subsection*{Heuristic description of correlation functions} 
The staggered density correlation function along the tilt are fitted with a staggered exponential decay $c_i = c_0 (-1)^i \exp(-i/L_\parallel)$, with an amplitude $c_0$ and the correlation length $L_\parallel$ as free parameters. For the case of weak transverse tunneling  we have added a small background Gaussian function to account for an artifact from slight displacements of the clouds between measurements. The monotonic density correlation functions perpendicular to the tilt are fitted with an exponential $c_i = c_0 \exp(-i/L_\perp)$ for the case of weak transverse tunneling, and with a Gaussian $c_i = c_0 \exp(-i^2/L^2_\parallel)$ for the case of strong transverse tunneling.



%

\end{document}


\title{Supplementary Information for: \\Formation of spontaneous density-wave patterns in DC driven lattices}
\author{H. P. Zahn}
\affiliation{Institut f{\"{u}}r Laserphysik, Universit{\"{a}}t Hamburg, 22761 Hamburg, Germany}
\author{V. P. Singh}
\affiliation{Zentrum f{\"{u}}r Optische Quantentechnologien, Universit{\"{a}}t Hamburg, 22761 Hamburg, Germany}
\affiliation{Institut f{\"{u}}r Theoretische Physik, Leibniz Universit{\"{a}}t Hannover, Germany}
\author{M. N. Kosch}
\affiliation{Institut f{\"{u}}r Laserphysik, Universit{\"{a}}t Hamburg, 22761 Hamburg, Germany}
\author{L. Asteria}
\affiliation{Institut f{\"{u}}r Laserphysik, Universit{\"{a}}t Hamburg, 22761 Hamburg, Germany}
\author{L.~Freystatzky}
\affiliation{Zentrum f{\"{u}}r Optische Quantentechnologien, Universit{\"{a}}t Hamburg, 22761 Hamburg, Germany}
\affiliation{The Hamburg Centre for Ultrafast Imaging, 22761 Hamburg, Germany}
\author{K. Sengstock}
\affiliation{Institut f{\"{u}}r Laserphysik, Universit{\"{a}}t Hamburg, 22761 Hamburg, Germany}
\affiliation{Zentrum f{\"{u}}r Optische Quantentechnologien, Universit{\"{a}}t Hamburg, 22761 Hamburg, Germany}
\affiliation{The Hamburg Centre for Ultrafast Imaging, 22761 Hamburg, Germany}
\author{L. Mathey}
\affiliation{Institut f{\"{u}}r Laserphysik, Universit{\"{a}}t Hamburg, 22761 Hamburg, Germany}
\affiliation{Zentrum f{\"{u}}r Optische Quantentechnologien, Universit{\"{a}}t Hamburg, 22761 Hamburg, Germany}
\affiliation{The Hamburg Centre for Ultrafast Imaging, 22761 Hamburg, Germany}
\author{C. Weitenberg}
\email{christof.weitenberg@physnet.uni-hamburg.de}
\affiliation{Institut f{\"{u}}r Laserphysik, Universit{\"{a}}t Hamburg, 22761 Hamburg, Germany}
\affiliation{The Hamburg Centre for Ultrafast Imaging, 22761 Hamburg, Germany}

\maketitle

\section{Derivation of the effective Hamiltonian}

We describe the 2D triangular lattice by the Bose-Hubbard Hamiltonian %
\begin{equation}
	\hat{H}_0=-J\sum_{\langle jk\rangle}\left( \hat{b}_j^\dagger \hat{b}_k + \hat{b}_k^\dagger  \hat{b}_j\right) + \frac{U}{2}\sum_j \hat{n}_j(\hat{n}_j-1),
\end{equation}
where $\hat{b}_j$ is the bosonic annihilation operator and $\hat{n}_j=\hat{b}_j^\dagger \hat{b}_j$ is the number operator at site $j$. $\langle jk\rangle$ denotes nearest neighbours. We add the lattice tilt perpendicular to a primitive lattice vector by the term $\hat{H}_\DD= \DD \sum_j x_j \hat{n}_j$, where $x_j$ is the $x$ coordinate of site $j$. 
The tilt energy $\DD$ is the dominant energy scale, since $\DD \gg J, U$. To include the effect of the tilt in the system, we go to the interaction picture with respect to $\hat{H}_\DD$ and obtain
\begin{align}
	\hat{H}_I = &\exp(i\hat{H}_\DD t)\hat{H}_0\exp(-i\hat{H}_\DD t), \\
	\hat{b}_j(t)=&\exp(i\hat{H}_\DD t)\hat{b}_j\exp(-i\hat{H}_\DD t)=\hat{b}_j\exp(-ix_j\DD t), \\
	\hat{b}_j^\dagger(t)=&\exp(i\hat{H}_\DD t)\hat{b}_j^\dagger\exp(-i\hat{H}_\DD t)=\hat{b}_j^\dagger\exp(ix_j\DD t).
\end{align}
This results in
\begin{equation}\label{eq:Hint}
	\hat{H}_I=\hat{H}_U+\hat{H}_J+\hat{H}_1\exp(-i\DD t)+\hat{H}_{-1}\exp(i\DD t),
\end{equation}
with
\begin{align}
	\hat{H}_U=&\frac{U}{2}\sum_{j}\hat{n}_j(\hat{n}_j-1), \\
	\hat{H}_J=&-J\sum_{\langle jk\rangle_y}\hat{b}_j^\dagger \hat{b}_k + \hat{b}_k^\dagger \hat{b}_j, \\
	\hat{H}_1=&-J\sum_{\langle jk\rangle_{x+}}\hat{b}_j^\dagger \hat{b}_{k}, \\
	\hat{H}_{-1}=&-J\sum_{\langle jk\rangle_{x-}}\hat{b}_{j}^\dagger \hat{b}_{k}.
\end{align}
	\begin{figure}[t]
		\includegraphics[width=0.95\linewidth]{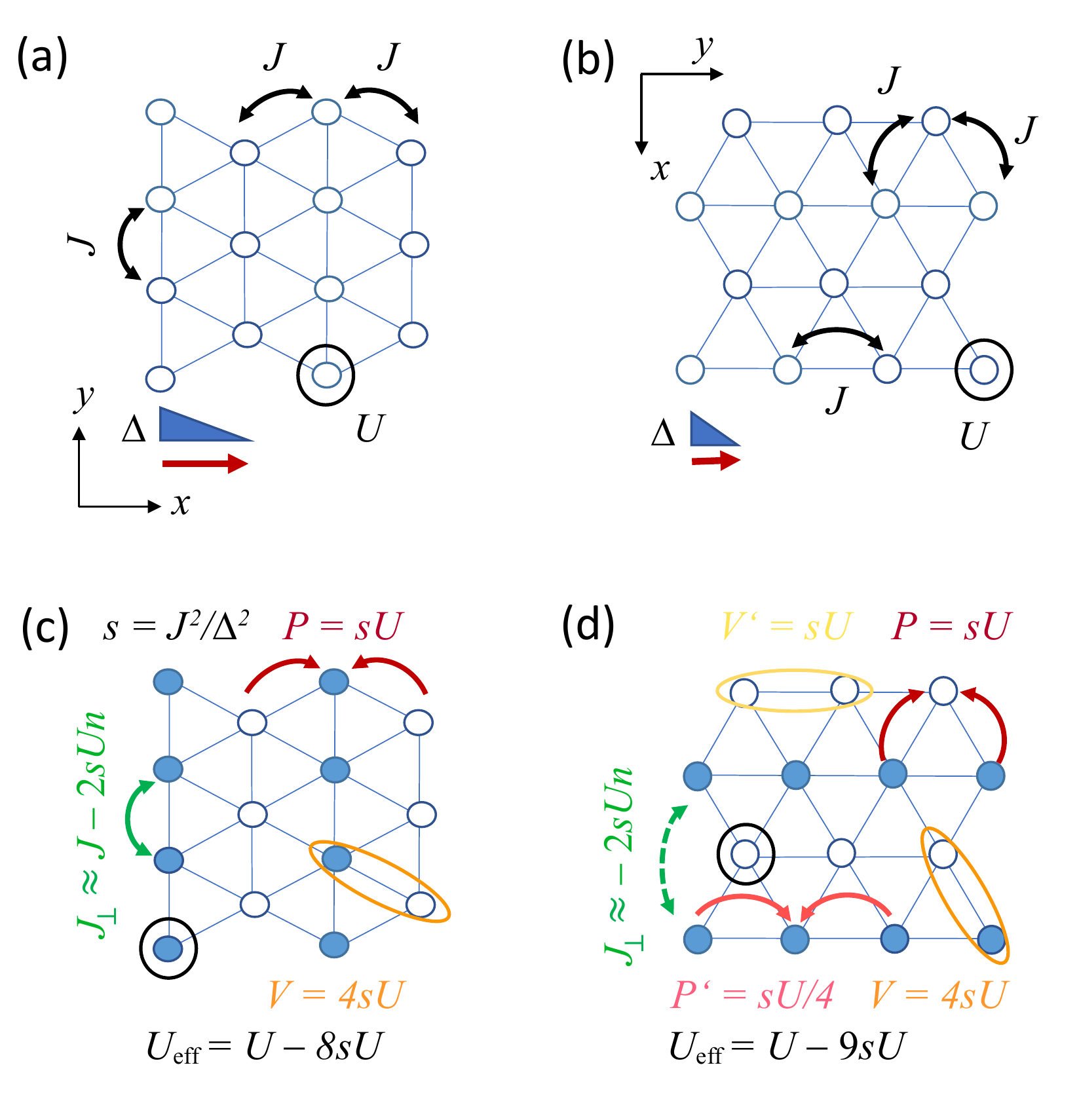}
		\caption{Effective Hamiltonian for the tilted lattice. Visualization of the tight binding model of the tilted lattice with energy offset $\Delta$, on-site interaction $U$ and direct hopping $J$ (a,b) and of the respective effective models including nearest neighbor interactions $V$ and pair tunneling $P$ as well as transverse tunneling $J_\perp$ and effective on-site interaction energy $U_{\rm eff}$ (c,d). Compared to the usual tilt direction (a,c), the perpendicular tilt direction (b,d) gives rise to a smaller perpendicular hopping $J_\perp$ as well as additional terms arising from resonances across two lattice sites (the primed quantities $V'$ and $P'$). The expression for the perpendicular hopping $J_{\perp}$ in mean-field approximation contains the mean on-site occupation $n$.} \label{fig:S1_EffectiveHamiltonian}
	\end{figure} 	
$\langle jk\rangle_y$ denotes nearest neighbours in the $y$ direction, $\langle jk\rangle_{x\pm}$ are the nearest neighbours in $x$ directions such that the tilt potential on site $k$ is bigger (smaller) than at site $j$.
Eq.~(\ref{eq:Hint}) is the form of a periodically driven system with frequency $\DD$, which  allows us to use a Magnus expansion \cite{Magnus1954,Blanes2009,Zhu2016}. 
We consider Magnus expansion to second order
\begin{align}
	\hat{H}_{\rm eff}^{(2,2)}&=-\frac{1}{2\DD^2} \big[ [\hat{H}_U+\hat{H}_J,\hat{H}_1],\hat{H}_{-1} \big] \nonumber \\
	& \quad -\frac{1}{2\DD^2} \big[ [\hat{H}_U+\hat{H}_J,\hat{H}_{-1}],\hat{H}_{1} \big]. 
\end{align}

We note that the first-order term vanishes in the expansion. We solve the double commutators and obtain the effective Hamiltonian

\begin{align}
\hat{H}_{\rm eff} &= \hat{H}_U+\hat{H}_J+\hat{H}_{\rm eff}^{(2,2)}  \\
& \equiv \hat{H}_{U,{\rm eff}} + \hat{H}_{J_\perp} + \hat{H}_{P},  
\end{align}
where
\begin{align}
	\hat{H}_{U,{\rm eff}}=&\left(\frac{U}{2}-4sU\right)\sum_j\hat{n}_j(\hat{n}_j-1)+4sU\sum_{\langle jk\rangle_y}\hat{n}_j\hat{n}_k, \\
	\hat{H}_P=&-sU\sum_{\langle jkl\rangle_\times} \hat{b}_j^\dagger \hat{b}_j^\dagger \hat{b}_k \hat{b}_l+\hat{b}_l^\dagger \hat{b}_k^\dagger \hat{b}_j \hat{b}_j, \\
	\hat{H}_{J_\perp}=&-\sum_{\langle jk\rangle_y}(J\hat{b}_j^\dagger \hat{b}_k + sU\hat{b}_j^\dagger \hat{n}_j \hat{b}_k + sU\hat{b}_j^\dagger \hat{n}_k \hat{b}_k +h.c.)\nonumber\\
	&+2sU\sum_{\langle jkl\rangle_\triangle}(\hat{b}_j^\dagger \hat{n}_l \hat{b}_k + \hat{b}_k^\dagger \hat{n}_l \hat{b}_j). \label{eq:dens-ind-tunnel}
\end{align}
$ \langle jk \rangle_{x(y)}$ denotes nearest neighbours in the $x(y)$ direction, $\langle jkl\rangle_\times$ are all pairs of bonds with one bond up and one down the tilt, and $\langle jkl\rangle_\triangle$ denotes triangular plaquettes such that $j$ and $k$ are nearest neighbors in the $y$ direction.

\begin{figure}[t]
	\includegraphics[width=0.95\linewidth]{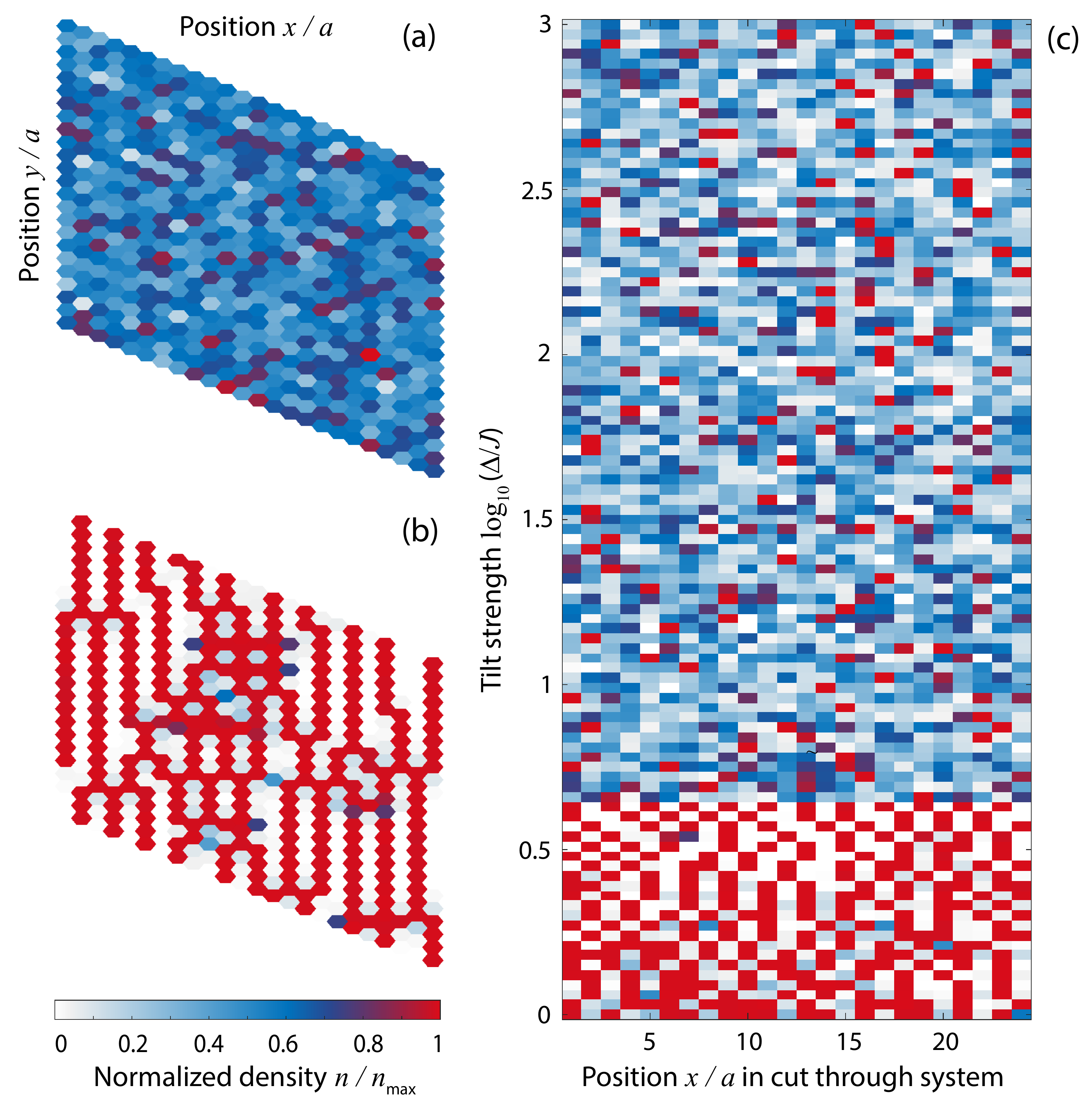}
	\caption{Monte Carlo simulations. (a),(b), Density distributions $n(x,y)$ of single Monte Carlo samples of the effective model for tilt energy $\Delta$ and tunnel energy $J$ with $\log_{10}(\Delta/J)=2.0$ (a) and $\log_{10}(\Delta/J)=0.5$ (b). The density is normalized by the maximum density $n_\mathrm{max}$ for each image. (c) Cuts through images as in (a),(b) as a function of $\Delta/J$. The density wave pattern is visible below a critical value of $\Delta/J  \sim 4.4$ and above this critical tilt the density is randomly fluctuating.} \label{fig:monte-carlo} 
\end{figure} 

Compared to the original model, most lattice parameters are slightly renormalised by the tilt. 
The interaction scales as $U\rightarrow U-8sU$. 
The hopping transverse to the tilt scales as $J_\perp \rightarrow J-2sUn$, where this correction comes from second order tunneling processes and $n$ is the mean-field density. The single-particle hopping along the tilt does not appear in the effective Hamiltonian. 
In $\hat{H}_{U,{\rm eff}}$ and $\hat{H}_P$, the new terms are nearest neighbour repulsion and pair hopping. 
The second term in Eq. (\ref{eq:dens-ind-tunnel}) is a density induced tunneling term, which appears transverse to the tilt \cite{Jurgensen2012}.

For a tilt along one lattice vector (Fig.~\ref{fig:S1_EffectiveHamiltonian}b,d) the general idea stays the same. In this case Eq.~(\ref{eq:Hint}) contains additional terms with $\exp(\pm 2i\DD t)$ and therefore additional terms appear in the effective Hamiltonian. On the other hand the direct hopping perpendicular to the tilt does not exist and coherence in this direction can only be built by second order processes.
To describe the effective Hamiltonian for this case, we define the tilt energy offset $\DD$ as indicated in Fig. \ref{fig:S1_EffectiveHamiltonian}. We obtain the effective Hamiltonian
\begin{align}
    \hat{H}_{\parallel, \rm eff} =& \hat{H}'_{U, {\rm eff}}+\hat{H}'_P+\hat{H}'_{J_\perp}+\hat{H}_{2\DD}, 
\end{align}  
with
\begin{align}
    \hat{H}'_{U,{\rm eff}}=&\left(\frac{U}{2}-\frac{9}{2}sU\right)\sum_j\hat{n}_j(\hat{n}_j-1)+4sU\sum_{\langle jk\rangle_x}\hat{n}_j\hat{n}_k,\\
    \hat{H}'_P=&-sU\sum_{\langle jkl\rangle_\times} \hat{b}_j^\dagger \hat{b}_j^\dagger \hat{b}_k \hat{b}_l+\hat{b}_l^\dagger \hat{b}_k^\dagger \hat{b}_j \hat{b}_j,\\
    \hat{H}'_{J_\perp}=&-sU\sum_{\langle jklm\rangle_\lozenge}\left(\hat{b}_j^\dagger \hat{n}_j \hat{b}_l + \hat{b}_j^\dagger \hat{n}_l \hat{b}_l\right. \nonumber\\ 
    &\quad\quad\quad\quad\quad\left. -2\hat{b}_j^\dagger \hat{n}_k \hat{b}_l - 2\hat{b}_j^\dagger \hat{n}_m \hat{b}_l+h.c.\right),\\
	\hat{H}_{2\DD}=&sU\sum_{\langle jk\rangle_{y}}\hat{n}_j\hat{n}_k - \frac{1}{4}sU\sum_{\langle jkl\rangle_{y}} \hat{b}_j^\dagger \hat{b}_j^\dagger \hat{b}_k \hat{b}_l+\hat{b}_l^\dagger \hat{b}_k^\dagger \hat{b}_j \hat{b}_j,
\end{align}
where $s=J^2/\DD^2$ has the same form as before.
$\langle jk\rangle_x$ are nearest neighbours in $x$ direction. Due to the geometry of the lattice these have an energy difference of $\DD$. $\langle jkl\rangle_\times$ are pairs of bonds with one bond up and one down the tilt and $\langle jklm\rangle_\lozenge$ denotes diamond shaped plaquettes such that sites $j$ and $l$ are at the same tilt energy.
$\langle jk\rangle_y$ are nearest neighbours in $y$ direction and $\langle jkl\rangle_y$ are pairs of bonds in $y$ direction. The interaction rescales as $U\rightarrow U-9sU$. Effective single particle hopping perpendicular to the tilt is present with strength $J_\perp=-2sUn$, where n is the mean-field density. In this case the hopping is to next nearest neighbours.
	
\begin{figure*}[t]
		\includegraphics[width=0.85\linewidth]{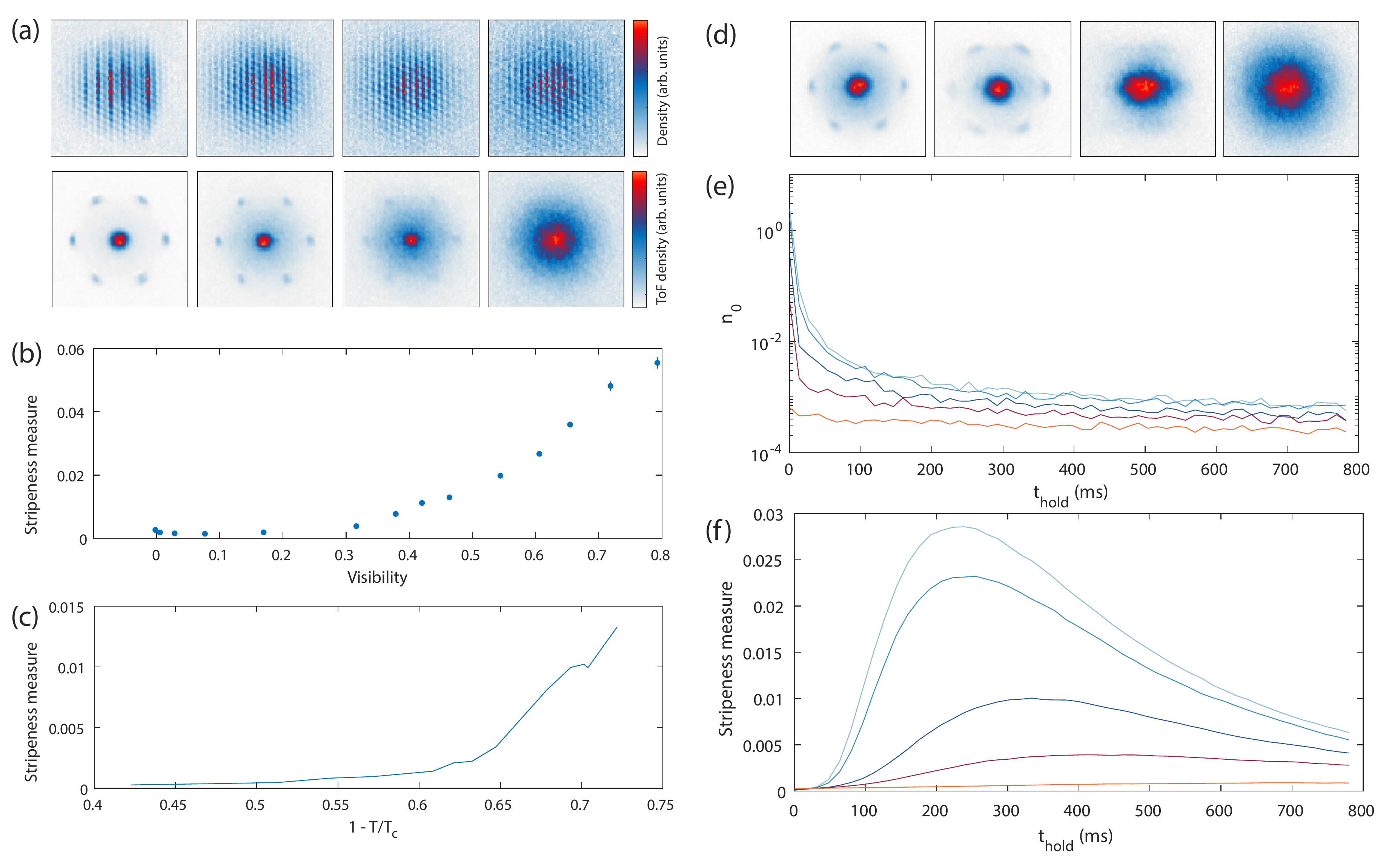}
		\caption{Influence of coherence on the density-wave formation. (a) Density wave pattern for different initial temperatures of the BEC before the quench into the tilted system (upper row). The coherence of the initial system is evidenced by the visibility of the Bragg peaks in a time-of-flight (ToF) measurement before the quench (lower row). The temperature is increased deliberately by a hold time in the lattice of (from left to right) (0.1, 750, 1200, 1800)\,ms leading to a visibility of (0.79, 0.42, 0.17, 0.006). (b) Density-wave order as a function of the Bragg-peak visibility in the experiment for a hold time of 100\,ms and an energy offset of $\Delta = h \times 1.4$\,kHz. All error bars correspond to the $68\%$ confidence interval. (c) Simulation of the density-wave order for the experimental parameters, plotted as a function of $1-T/T_{\rm c}$ where $T_{\rm c}$ is the critical temperature of a non-interacting gas. 
		(d) ToF measurements for (0.1, 1.1, 2.1, 1200)\,ms (from left to right) hold time after the quench and $\Delta = h \times 1.1$\,kHz. The hold times are approximately stroboscopic with respect to the Bloch oscillation period. (e) The condensate density $n_0$ versus time in the numerical simulation for different temperatures reproduces the rapid loss of coherence and shows that a very small coherent fraction remains, which is not experimentally detectable. (f) The density-wave order for the parameters in (e) shows that this residual coherence is crucial for the density wave formation. The different colours stand for different initial temperatures of $T/T_c= 0.28$, $0.32$, $0.39$, $0.45$, and $0.53$ (top to bottom).} \label{fig:S3_Coherence}
	\end{figure*} 	

\section{Short time dynamics}

To understand the emergence of the density wave we consider a 1D lattice system. The initial dynamics are dominated by the fast phase rotation with frequency $\DD$. The system is described by the Hamiltonian (see Eq.~(\ref{eq:Hint}))
\begin{align}
\hat{\cH} &= -J\sum_j\left(\hat{b}_j^\dagger \hat{b}_{j+1}e^{-i\DD t}+\hat{b}_{j+1}^\dagger \hat{b}_{j}e^{i\DD t}\right)\nonumber\\
&\quad+\frac{U}{2}\sum_j\hat{n}_j(\hat{n}_j-1).
\end{align}
To analyze the density wave order we calculate the density-density correlation function $\langle \hat{n}_{j_1}\hat{n}_{j_2}\rangle$. Within a perturbative expansion, for $\hat{O}=\hat{n}_{j_1}\hat{n}_{j_2}$ we have 
\begin{align}
	\hat{O}(t)=&\hat{O}+i\int_{0}^{t}\text{d}t_1[\hat{\cH}(t_1),\hat{O}]\nonumber\\
	&-\int_{0}^{t}\text{d}t_1\int_{0}^{t_1}\text{d}t_2[\hat{\cH}(t_2),[\hat{\cH}(t_1),\hat{O}]],\label{eq:pert}
\end{align}
where we considered expansion to second order. 
We solve the commutators and then take the expectation value. The first-order term drops out and the second-order term has the form
\begin{align}
	&\langle \hat{b}_{j_1}^\dagger \hat{b}_{j_2} \hat{b}_{j_3}^\dagger \hat{b}_{j_4}\rangle  \approx \nonumber \\
	&\quad \frac{g(|j_1-j_2|)g(|j_3-j_4|)g(|j_1-j_4|)g(|j_2-j_3|)}{g(|j_1-j_3|)g(|j_2-j_4|)}.
\end{align}
$g(j)=\langle \hat{b}_{0}^\dagger \hat{b}_{j}\rangle \approx n_0 \exp(-|j|/r_0)$, where $n_0$ is the equilibrium particle number and $r_0$ is the correlation length.
We calculate the time averaged correlations $\overline{\langle \hat{n}_{j_1}\hat{n}_{j_2}\rangle}$ by keeping the constant terms in Eq.~(\ref{eq:pert}) and obtain
\begin{align}
	\overline{\langle \hat{n}_{j_1}\hat{n}_{j_2}\rangle}
	&=g^2(0)+sg^2(1)\left(8F(R)-\frac{4}{F(R)}-4F(R+1)\right.\nonumber\\
	\quad &\left.+\frac{2}{F(R+1)}-4F(R-1)+\frac{2}{F(R-1)}\right),
\end{align}
where we used $R=a|j_1-j_2|$ as the distance between sites with the lattice spacing $a$ and $s=J^2/\DD^2$.
We define 
\begin{equation}
	F(R)=\frac{g^2(R)}{g(R+1)g(R-1)}.
\end{equation}
We have $F(R)=\delta_{R,0}(g^2(0)/g^2(1)-1)+1$ and $1/F(R)=\delta_{R,0}(g^2(1)/g^2(0)-1)+1$.
In momentum space the correlations are
\begin{align}
	\langle \hat{\rr}_{k}\hat{\rr}_{k}\rangle&=\sum_R\exp(ikR)\overline{\langle \hat{n}_{j_1}\hat{n}_{j_2}\rangle}\\
	&=4sn_0^2(1-\cos(ka))(2-e^{-2a/r_0}-e^{-4a/r_0}).
\end{align}
This yields a density-density correlation peak at $k=\pi/ a$, corresponding to the density pattern of period $2$. The magnitude of the correlations scales with $s$ and thus $\DD^2\langle \rr_{k}\rr_{k}\rangle$ is independent of $\DD$, as we describe in Fig.~3c of the main text.
	
\section{Monte Carlo Simulation of the effective model}
We perform Monte Carlo simulations to determine the equilibrium states of the 2D effective model for a wide range of the tilt strength. 
For convergence of Monte Carlo simulation, we consider a homogeneous density and a triangular lattice of size  $24\times 24$. 
We use the same $J= h \times 13\,$Hz and the same $U/J =0.18$ as in the experiments and the temperature $T/J = 25$.
We vary the energy offset $\Delta$ in the range $\Delta/J=2-1000$ and sample the corresponding initial state via classical Metropolis Monte Carlo as in the c-field simulations. 
For each $\Delta$ we carry out $1$ million Monte Carlo steps per site to obtain the equilibrium state. 
In Fig.~\ref{fig:monte-carlo}, we show the density along the $y$ direction of the equilibrium state of a single sample. 
The density displays a density-wave pattern for $\Delta$ up to a critical value of $\Delta/J  \sim 4.4$. 
Above this critical value, no density wave is visible and we only observe a fluctuating density of the sample. 
This is in contrast to our experiments, where a density wave appears as a transient state for the tilt energies that are about 100 times larger than the effective model.   
Such small values of $\Delta/J$ are hard to realize via the effective model of a tilted system, because the derivation relies on large $\Delta/J \gg 1$.

\section{Influence of coherence}	
We study the importance of coherence for the formation of the density-wave order. As a first point, we compare the density-wave order for different initial temperatures before the quench into the tilted system and find that the density-wave order vanishes when no initial coherence is present (see Fig.~\ref{fig:S3_Coherence}a-c). 

As a second point, we investigate the coherence after the quench and find that while most coherence is rapidly lost after a few Bloch oscillation periods (Fig.~\ref{fig:S3_Coherence}d), a residual coherence is detectable in the simulation results and the density-wave formation vanishes as a function of temperature together with this residual coherence (Fig.~\ref{fig:S3_Coherence}e,f). The simulated condensate density $n_0$, which is a measure of the coherence between the lattice sites, is determined by averaging the complex field $\psi_i$ within a central region with a radius of 6 sites and then averaging the absolute value squared over the thermal ensemble, i.e., $n_0 = \langle |1/N_{\rm roi}\sum_{i} \psi_i |^2 \rangle_{\rm ensemble}$ with $N_{\rm roi}$ being the number of sites in the region of interest. The data in Fig.~\ref{fig:S3_Coherence}d are for the tilt perpendicular to a lattice vector, where the residual coherence also shows in the experimental data as a small anisotropy of the time-of-flight distribution. For the other tilt direction with the much smaller tunneling along the columns, the coherence is smaller and we do not observe an anisotropy in the time-of-flight images. Due to the small coherent fraction, our density wave would not be visible as a second Bragg peak in time-of-flight measurements as in refs.~\cite{Fallani2004,Cristiani2004,Gemelke2005} and therefore explicitly requires the novel microscopy technique of ref.~\cite{Asteria2021}.
		
	\begin{figure}[t]
		\includegraphics[width=0.95\linewidth]{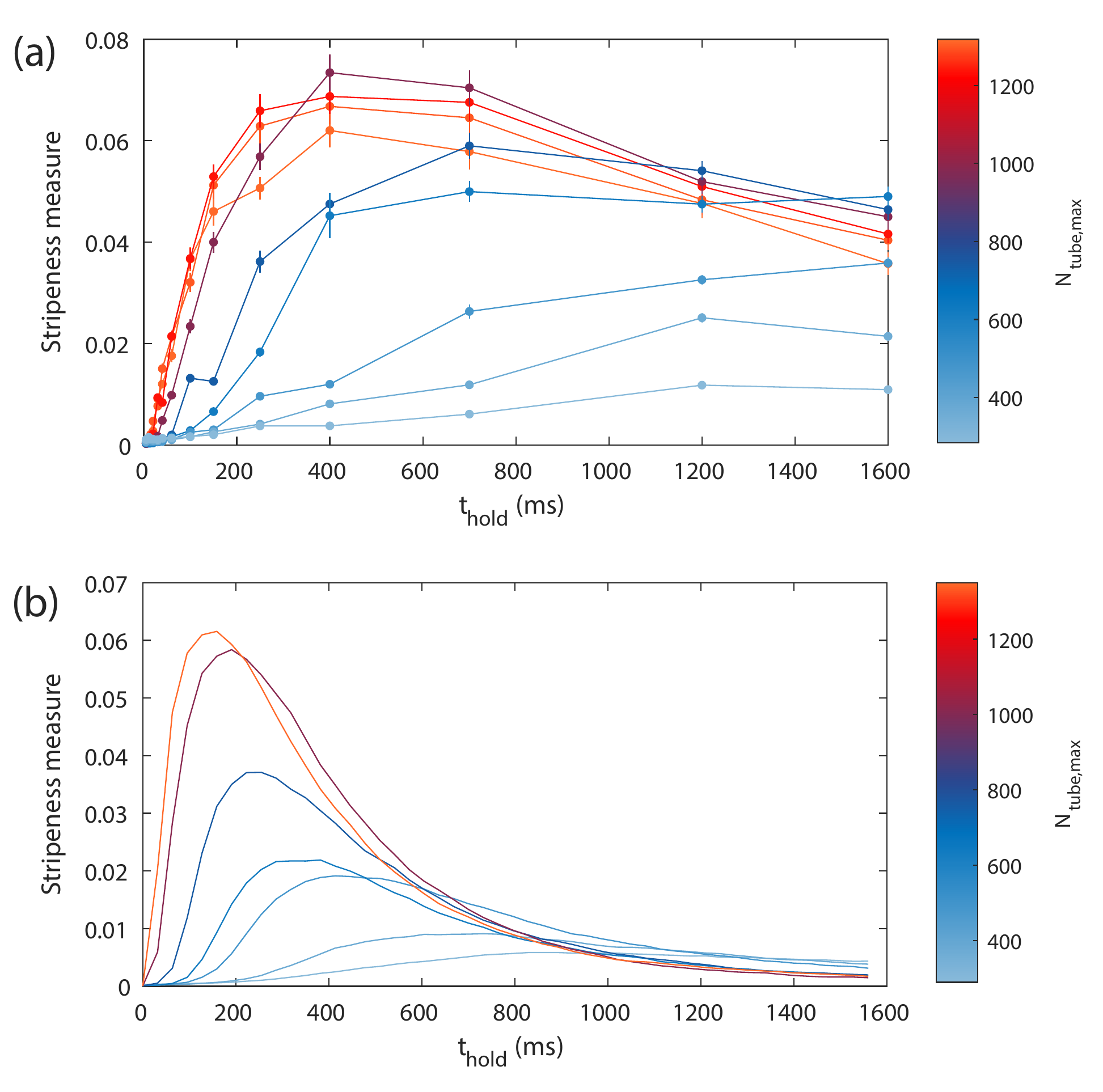}
		\caption{Atom number dependence. (a) Experimental density wave order dynamics as a function of the initial atom number (color coded using the biggest tube population at $t \sim 0$). The energy offset was $\Delta = h \times 1.4$\,kHz. All error bars correspond to the $68\%$ confidence interval.
		(b) Simulation for the corresponding parameters and $k_{\rm B} T/J =100$, without including atom loss.
		} \label{fig:S4_Atom-number}
	\end{figure} 
		
	\begin{figure}[t]
		\includegraphics[width=0.9\linewidth]{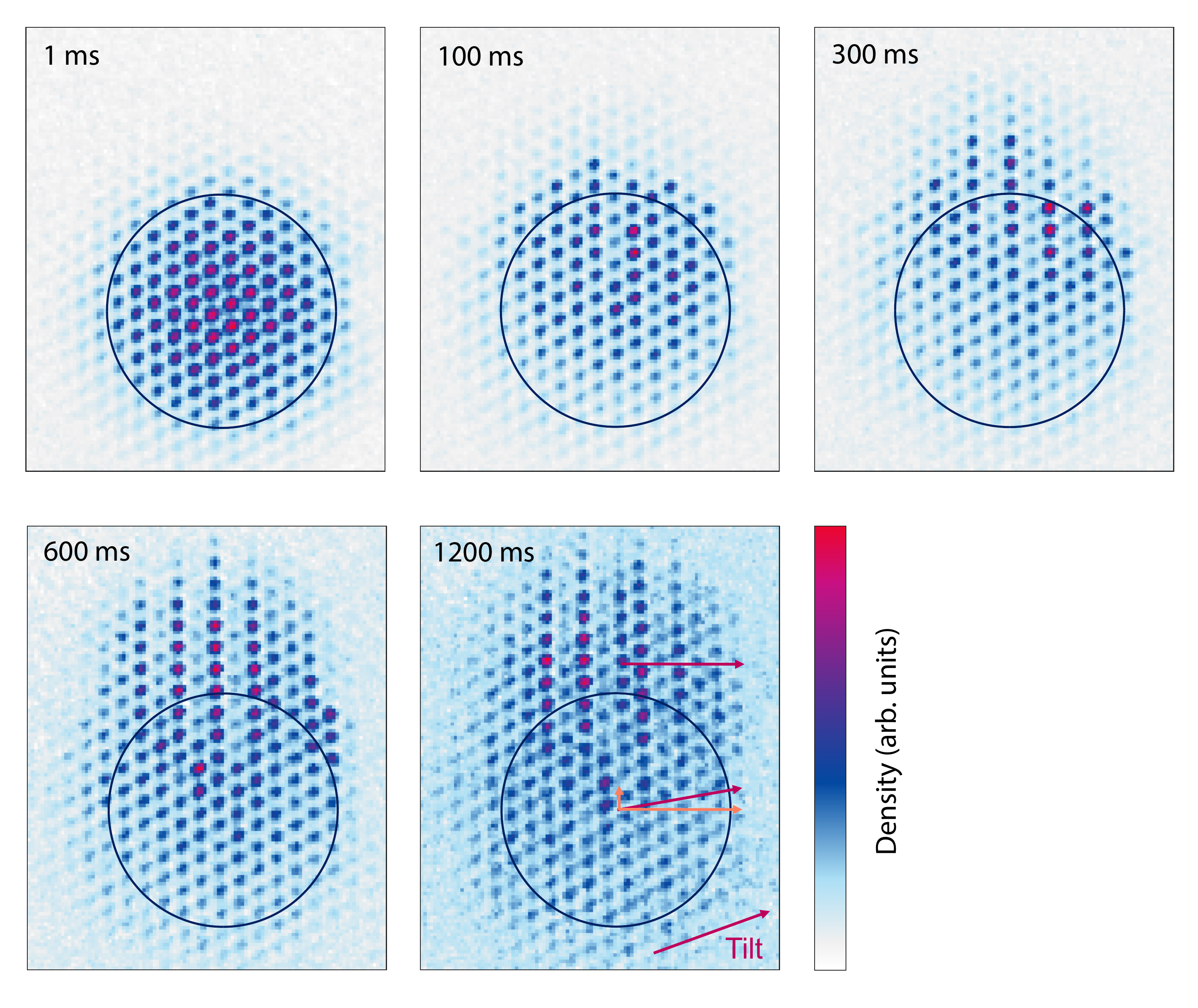}
		\caption{Incommensurate tilt direction. Experimental images for different hold times after quenching to a tilt, which is 10° tilted against the direction perpendicular to a direct lattice vector. The red arrows indicate the local tilt direction, the orange arrows show the projection of the tilt onto a lattice vector and onto the direction perpendicular to it. A fraction of the cloud moves from the initial position (circles) to a position where the tilt is again commensurate (topmost arrow in the last subfigure) and forms a density wave with large contrast.} \label{fig:S4_other-tilt-direction}
	\end{figure} 	
    
    \begin{figure}[t]
		\includegraphics[width=0.95\linewidth]{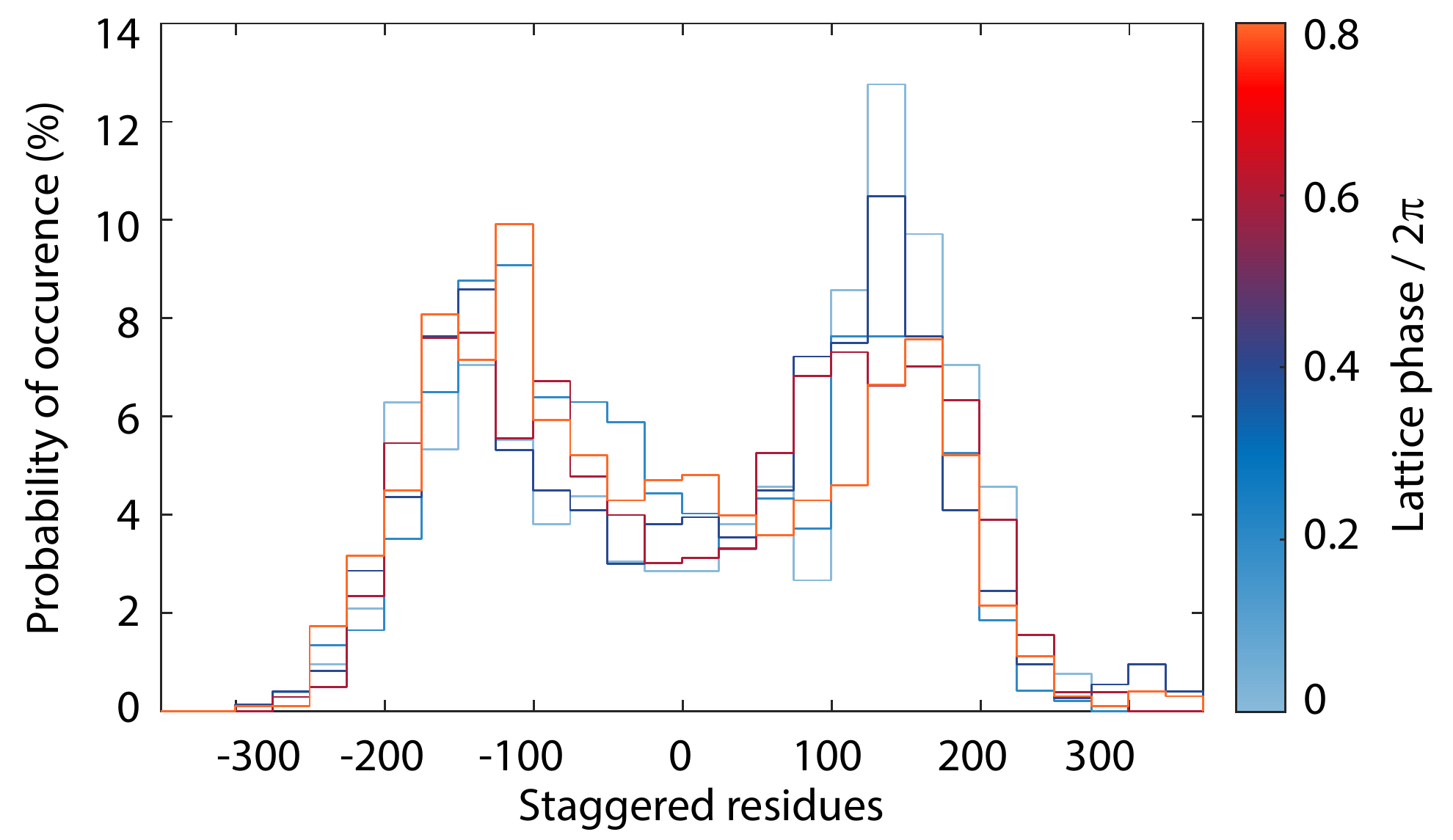}
		\caption{Lattice-phase resolved histogram. The histogram of local density-wave order of Fig.~2c of the main text with separate analysis for the different lattice phases (color coded) still shows the double peak structure reflecting the spontaneous symmetry breaking.} \label{fig:S5_lattice-phase-res-histo}
	\end{figure} 
	
\section{Confirmation of the interaction-driven dynamics}
In Fig.~3 of the main text, we have observed the interaction-driven nature of the density-wave formation via the slowing down of the long-time dynamics due to atom losses. Here, we further strengthen this point, by studying the formation dynamics of the density wave for different initial atom numbers (Fig.~\ref{fig:S4_Atom-number}). Indeed, we find much faster build-up of the density-wave order with increasing atom number both in the experiment and the numerical simulation.

\section{Incommensurate tilt directions}
We also study incommensurate tilt directions, which are not along or perpendicular to a direct lattice vector, by shifting the magnetic trap in the appropriate directions. The dynamics for a tilt direction 10$^\circ$ rotated against the direction perpendicular to a direct lattice vector is shown in Fig.~\ref{fig:S4_other-tilt-direction}. Because the tilt is produced by the shifted magnetic trap, its direction changes with position. The force from the tilt can be decomposed into a component perpendicular to a lattice vector, where the atoms are Stark localized, and a component along a lattice vector, where the atoms can move. Indeed, a fraction of the atoms moves along the latter direction. For the given geometry, the component along the lattice vector vanishes after the atoms have moved by about 5 lattice sites and the motion stops there. The atoms automatically align with a commensurate tilt direction.	

This splitting is reminiscent of early experiments in Josephson junction arrays finding that the superfluid fraction can move through the lattice, while the thermal fraction is pinned \cite{Cataliotti2001, Cataliotti2003}. The observed separation might therefore point a way to distill the zero-temperature part of the cloud. The formation of the density-wave is restricted to a small region with minimal tilt along the stripe pattern and reaches a larger contrast than in the commensurate situations discussed in the main text. This is in line with the observed relevance of coherence discussed above.  

\section{Postselection of lattice-phase drifts}
The lattice phase, i.e. the exact position of the optical lattice below a lattice site relative to the initial cloud center, is stable during one experimental preparation, but random between different experimental images. The high-resolution imaging allows to read out this phase and to postselect on it \cite{Asteria2021}. For the investigation of the spontaneous symmetry breaking, it is important to verify that the apparently random position of the density wave is not simply triggered by this lattice phase. To do so, we repeat the analysis of the domains in Fig.~2c after postselection on the lattice phase and present lattice-phase resolved histograms in Fig.~\ref{fig:S5_lattice-phase-res-histo}. While the statistics of the individual histograms is less extensive now, one can still recognize the double peak structure in all histograms, which confirms that the symmetry breaking of the density-wave is spontaneous and not triggered by the random lattice phase. 

\section{Pinning of the density-wave pattern for long hold times}
\begin{figure}[t]
		\includegraphics[width=0.95\linewidth]{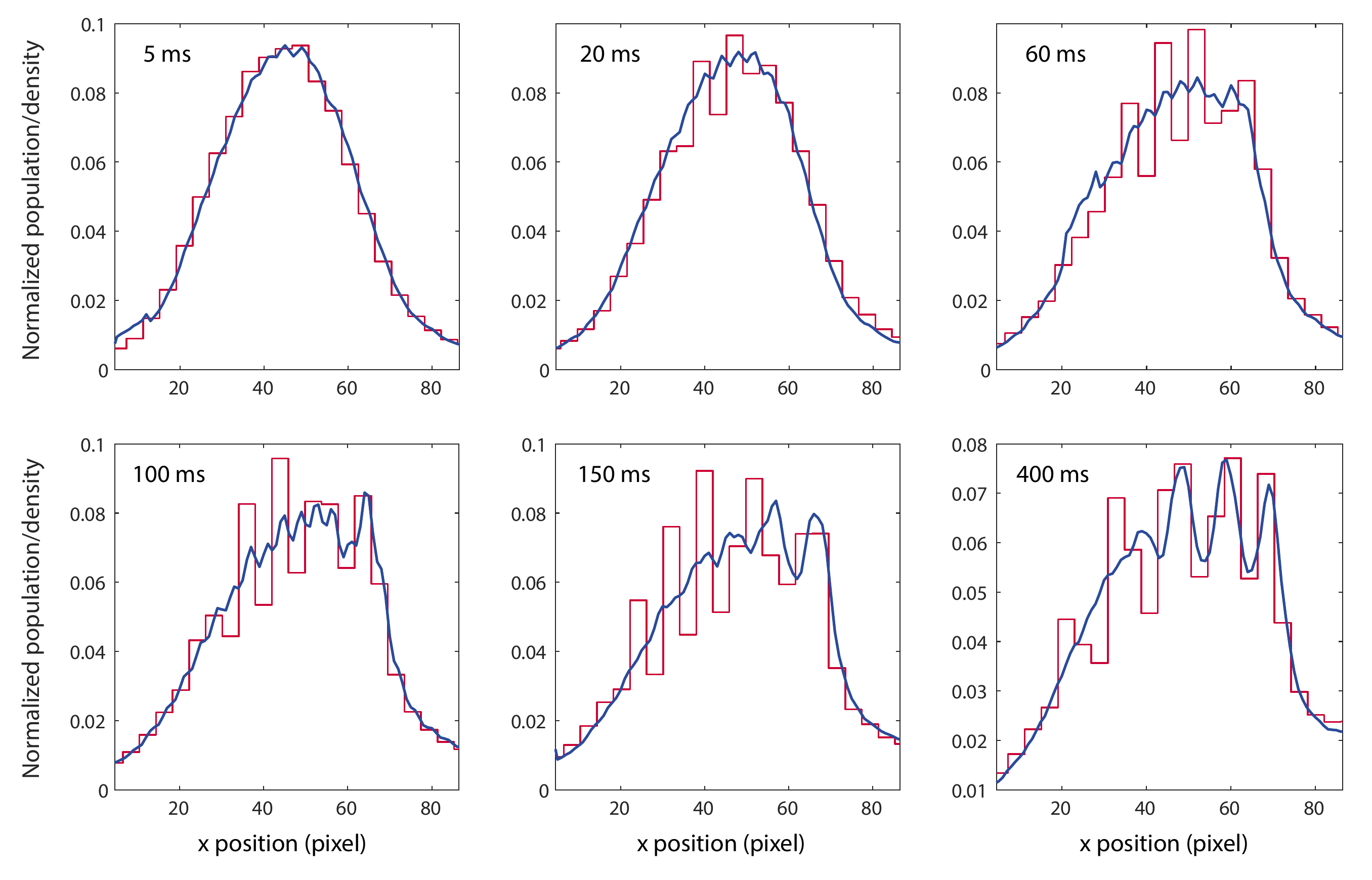}
		\caption{Spontaneous symmetry breaking and pinning. Long-term dynamics of the density waves illustrated by the populations integrated along the columns. Exemplary single shot profiles (red line) show the density wave formation while the absence of a density wave in the average profile (blue line) for early times confirms the spontaneous nature of the symmetry breaking. The average profile is continuous due to the varying lattice phases of the individual images. For later times, the average profile develops density waves as well, indicating the onset of pinning, which we attribute to self-trapping.
		}
        \label{fig:pinning}
	\end{figure}

For longer hold times (Fig.~\ref{fig:pinning}), we find an intriguing interplay of pair-tunneling and single-particle tunneling: the average profile develops density waves as well, which we attribute to steep density gradients formed due to self-trapping \cite{Anker2005}. This pins the density wave at a fixed position making the symmetry breaking not spontaneous anymore. We observe experimentally an inwards propagation of the pinning, starting from the edge at the lower end of the slope. This behavior seems reminiscent of the spontaneous formation and later stabilization also observed for supersolid droplets in continuous systems \cite{Sohmen2021}.

%


\setcounter{equation}{0}
\setcounter{figure}{0}
\setcounter{table}{0}
\renewcommand{\theequation}{S\arabic{equation}}
\renewcommand{\thefigure}{S\arabic{figure}}
\renewcommand{\bibnumfmt}[1]{[S#1]}
\renewcommand{\citenumfont}[1]{S#1}
